\documentclass[10pt,twocolumn,nofootinbib,aps,prd]{revtex4-1}
\usepackage{CJK}
\usepackage{amsmath,amssymb,bm,mathrsfs}
\usepackage{mathtools,mathptmx,array,slashed}
\usepackage[colorlinks=true,linkcolor=blue,citecolor=blue,urlcolor=blue]{hyperref}
\usepackage{color}
\usepackage{xcolor}
\usepackage{multirow}

\def\be{\begin{equation}}
\def\ee{\end{equation}}
\def\bea{\begin{eqnarray}}
\def\eea{\end{eqnarray}}
\def\ba{\begin{aligned}}
\def\ea{\end{aligned}}
\def\nn{\nonumber}
\def\p{\partial}
\def\cM{\mathcal{M}}
\def\cN{\mathcal{N}}
\def\cV{\mathcal{V}}
\def\cR{\mathcal{R}}
\def\cT{\mathcal{T}}

\begin{document}

\begin{CJK*}{GBK}{song}

\title{Static four-charge squashed black hole in five-dimensional $STU-W^2U$ supergravity
and its thermodynamics}

\author{Di Wu}
\email{wdcwnu@163.com}
\author{Shuang-Qing Wu}
\email{Contact author: sqwu@cwnu.edu.cn}

\affiliation{School of Physics and Astronomy, China West Normal University, Nanchong,
Sichuan 637002, People's Republic of China}

\date{\today}

\begin{abstract}
In this paper, we present a remarkably simple expression for the exact solution to the $D = 5$,
$\cN = 2$ supergravity coupled to three vector multiplets with the pre-potential $\mathcal{V}
= STU -W^2U \equiv 1$, which represents a five-dimensional static Kaluza-Klein black hole with
squashed $S^3$ horizons and four independent electric charges. It is asymptotically locally flat
and has a spatial infinity $R \times S^1 \hookrightarrow S^2$. We compute its conserved charges
via the counterterm method and demonstrate that the thermodynamic quantities satisfy both the
first law and Bekenstein-Smarr mass formula, provided the length of the compact extra dimension
is treated as a thermodynamic variable. \textcolor{black}{Besides, we have, for the first time,
observed a novel black hole splitting phenomenon induced by the extra charge -- namely, when the
fourth Abelian vector field is added, the black hole solution to the usual $STU$ model is split
into two different branches in the new $STU-W^2U$ model. This fact hints that the solution space
of the $STU-W^2U$ model has a much richer structure than that of the usual $STU$ model. We also
demonstrate that the special $S^2U-W^2U$ model can be converted into the usual $STU$ model which
suggests that one should coherently set $\rho_2 = 0$ in the general expression of the structure
function: $f(\rho) = 1 -\rho_1/\rho +\rho_2/\rho^2$.}
\end{abstract}

\maketitle

\end{CJK*}

\section{Introduction}

\textcolor{black}{Black holes in higher dimensions have a much richer and more complicated structure
than four-dimensional ones. In particular, of} various kinds of black objects in five dimensions,
there exists a class of Kaluza-Klein (KK) type black holes \cite{PTPS189-7} named as squashed black
holes after the initial work of Ishihara and Matsuno \cite{PTP116-417}, who adopted a specific measure
--- ``squashing transformation" to cast the five-dimensional asymptotically flat cohomogeneity-one
black hole solution with an $S^3$-spherical horizon into one \textcolor{black}{with the shape of} its
event horizon being a squashed three-sphere. The resulting five-dimensional spacetime is asymptotically
locally flat and has a spatial infinity $R \times S^1 \hookrightarrow S^2$, namely, its asymptotic
structures is identical to that of the Kaluza-Klein magnetic monopole --- a twisted $S^1$ fiber bundle
over a four-dimensional Minkowski spacetime at the spatial infinity. Under dimensional reduction along
the spatial direction, the squashed spacetime then becomes an exact black hole solution in the
four-dimensional KK theory. \textcolor{black}{Therefore, the five-dimensional Ishihara-Matsuno static
charged solution with squashed horizons can be thought of as describing an electrically charged black
hole on a Taub-NUT instanton, or a black hole that resides in the KK spacetime.}

The squashing algorithm first used in Ref. \cite{PTP116-417} is exceedingly simple, i.e.,
via multiplying some metric components by different powers of a squashing function to produce
new solutions within the same theory. Its extraordinary simplicity has attracted considerable
research interest over the past two decades, and prompted the derivation of numerous other
five-dimensional squashed black holes in various theories, including vacuum gravity \cite{NPB756-86},
Einstein-Maxwell-dilaton theory \cite{PRD74-024022,PRD85-064021,EPJC73-2377,PRD93-064052}, minimal
\cite{CQG25-245007,PRD78-064016,PRD78-124006,PTP121-823,PRD77-044040} and $U(1)^3$ \cite{1009.3568}
supergravity theories, etc. In particular, solutions in the background of a G\"{o}del universe were
considered in Refs. \cite{CQG25-245007,PRD78-064016,PRD78-124006,PTP121-823}. Research has since
expanded to encompass their thermodynamics \cite{PLB639-354,CQG24-4525,CQG25-085006,IJMPA24-2357,
PRD84-124040,PLB726-404,GRG48-154,EPJC77-706,NPB828-273}, Kerr/CFT correspondence \cite{NPB828-273},
Hawking radiation \cite{PRD76-064022,PRD77-024039,EPJC65-281,PRD83-064016}, quasinormal modes
\cite{PLB665-392,PRD79-084005}, strong gravitational lensing \cite{PRD81-124017,PRD83-124019,
JHEP0314089,ASS343-559,PLB728-170}, stability \cite{PRD77-064015,PRD77-084019,CQG27-215020},
geodesic motion \cite{PRD80-104037}, black hole shadows \cite{JHEP1019269}, etc. Incidently,
let us just only mention that there is also another kind of cohomogeneity-two KK black holes in
five-dimensional vacuum gravity, minimal and $U(1)^3$ supergravity theories \cite{NPB454-379,
NPB575-211,PRD79-064020,CQG26-125016,JHEP1110133,CQG26-145006,PRD84-024009,PRD86-024022,PRD87-024027}.

Although the squashing transformation offers a direct route to get new black hole solutions, it
often comes at the cost of extremely complicated expressions for the thermodynamic quantities,
since one has to make further coordinate transformations and certain troublesome identifications
of the solution parameters to render the appropriate asymptotic behavior of the metric at infinity.
What is more, because this procedure often leads to the non-vanishing of the modulus of the scalar
fields asymptotically at infinity, the first law generally acquires the contribution of the scalar
hairs \cite{PRL77-4992}.

By contrast, one can avoid this complexity via an alternative strategy in that one first tries to seek
a simpler expression for the solution which possesses the appropriate asymptotic behavior, then much
more elegant thermodynamic quantities can be cleanly computed. This methodology was previously applied
in Ref. \cite{PLB726-404} to find a new simple form for the static three-charge squashed black hole
in the five-dimensional $U(1)^3$ supergravity, which facilitates a thorough thermodynamic study. In
a subsequent work, this approach was extended to derive new forms of the five-dimensional neutral
rotating black hole \cite{GRG48-154}.

Despite these progresses, the construction of charged, squashed black holes in the five-dimensional
$\cN = 2$ supergravity with multiple vector multiplets, for instance, the $STU-W^2U$ model, remains
comparatively less explored. In this work, we shall focus on \textcolor{black}{this kind of static
squashed KK black hole solution within} the $STU-W^2U$ model of five-dimensional $\cN = 2$ supergravity
theory \cite{JHEP0226252,EPJC86-187}. This model constitutes a non-trivial extension of the well-known
$STU$ model by incorporating an additional vector multiplet that allows for a more intricate structure
of gauge interactions and scalar potentials. Recently, by directly employing the squashing transformation
\cite{PTP116-417} to the five-dimensional static black hole solution with four different electric charges
in the $STU-W^2U$ model, we succeed in constructing its static squashed counterpart, i.e, Eqs. (4.1) and
(4.2) in Ref. \cite{JHEP0226252}. The obtained solution represents a KK type black hole with its horizon
geometry being a squashed three-sphere, in other words, it is asymptotically locally flat and has a spatial
infinity $R\times S^1 \hookrightarrow S^2$. However, because the scalar moduli does not vanishes asymptotically
at infinity after further coordinate transformations and appropriate parameter identifications, the computed
expressions for the associated thermodynamic quantities are quite involved, and the first law also includes
the contribution of the scalar hairs \cite{PRL77-4992}.

On the contrary, one can use a brute-force method, as did in Ref. \cite{PLB726-404,GRG48-154}, to get a
relatively simple expression for the static squashed black hole solution in which three scalar fields
vanish asymptotically at infinity, rather than what was done in \cite{JHEP0226252} by directly applying
the squashing transformation to get its squashed version.\footnote{\textcolor{black}{Due to the presence of
the factor: $Z_1Z_2 -Z_4^2$, the suitable seed solution to which the squashing transformation is applied
should be the one that was solved by using the general harmonic functions: $Z_I = h_I +q_I/r^2$, rather
than adopting the specific choice: $h_I = (1, 1, 1, 0)$ as did in Ref. \cite{JHEP0226252}.}\label{fn1}}
In doing so, not only is the solution much simpler, but also the computed thermodynamic quantities are
very concise, facilitating the study of its thermodynamical property. In this paper, our aim is to first
find another new form for the four-charge static squashed black hole solution with which we can very easily
calculate its conserved mass and gravitational tension by using the counterterm method, and then we obviously
show that all thermodynamical quantities computed for our four-charge static squashed black hole perfectly
satisfy both the differential and integral mass formulae of usual black hole thermodynamics, provided
that the length of the compact extra dimension is regarded as a thermodynamical variable. \textcolor{black}{
In addition, we will uncover a novel black hole splitting phenomenon induced by the extra charge, which
had not been reported by any one before. Also, it is of some interest to study a special case ($q_2 = q_1$),
which belongs to the special $S^2U-W^2U$ model that can be attributed to the usual $STU$ model.}

The remaining part of our paper is organized as follows. In Sec. \ref{II}, we briefly review the
action of the $STU$ model of the $D = 5$, $\cN = 2$ supergravity and the three-charge static black
hole with a squashed horizon. In Sec. \ref{III}, we will first introduce the $STU-W^2U$ model with
the pre-potential $\cV = STU -W^2U \equiv 1$, and then present a simple expression for the four-charge
static squashed black hole solution. \textcolor{black}{In this section, after we concisely present
the main solving steps to arrive at our general non-extremal static four-charge squashed black hole
solution, we then present a novel black hole splitting phenomenon induced by the fourth Abelian
vector field -- with the addition of the fourth extra charge, the black hole solution to the usual
$STU$ model divides into two different branches in the new $STU-W^2U$ model. In particular in Sec.
\ref{IV}, we also show that the special $S^2U-W^2U$ model can be associated with the usual $STU$
model which hints that perhaps one should set $\rho_2 = 0$ in general.} In Sec. \ref{V}, the boundary
counterterm method is adopted to calculate the conserved mass and gravitational tension, which,
together with the entropy, horizon temperature, four charges, and their corresponding electrostatic
potentials, completely satisfy both the first law and Bekenstein-Smarr mass formula when the length
of the extra dimension is considered as a thermodynamical variable. Finally, Sec. \ref{VI} is
concluded with a summary and an outlook for future work. In the Appendix, we also present the
supersymmetric BPS solutions of the five-dimensional four-charge static squashed black hole.

\section{Three-charge static squashed black hole solution}\label{II}

In this section, we begin with a brief review of the five-dimensional $\cN = 2$, $U(1)^3$ supergravity
and the three-charge static squashed black hole solution presented in \cite{PLB726-404}. The complete
bosonic Lagrangian for $D = 5$, $\cN = 2$ ungauged supergravity coupled to two Abelian vector
multiplets for the $STU$ model reads:
\bea
S_5 &=& \frac{1}{16\pi G}\int d^{5}x \bigg\{\sqrt{-g}\Big[R -\frac{1}{2}(\p\phi_1)^2
 -\frac{1}{2}(\p\phi_2)^2  \nn \\
&& -\sum_{I=1}^3\frac{1}{4}X_I^{-2}F_{\mu\nu}^I F^{I\mu\nu}\Big] -\frac{1}{4}
 \varepsilon^{\mu\nu\alpha\beta\lambda}F_{\mu\nu}^1F_{\alpha\beta}^2A_{\lambda}^3\bigg\} \, .
\label{L-STU}
\eea
where the Chern-Simons term is included for the completeness, but it makes no contribution in
the nonrotating case. $R$ is the bulk Ricci scalar curvature, and $F_I = dA_I$ are the field
strength one-forms associated to three $U(1)$'s gauge fields. Two dilaton scalar fields $(\phi_1,
\phi_2)$ are related to three scalar functions $X_I$ by
\bea
X_1 = e^{-\phi_1/\sqrt{6} -\phi_2/\sqrt{2}} \, , \quad
X_2 = e^{-\phi_1/\sqrt{6} +\phi_2/\sqrt{2}} \, , \quad
X_3 = e^{2\phi_1/\sqrt{6}} \, . \nn
\eea

For the five-dimensional static squashed black hole, its asymptotic geometry approaches to
\be
ds^2 \simeq -d\tau^2 +d\rho^2 +\rho^2\big(d\theta^2 +\sin^2\theta\, d\phi^2\big)
 +L_{\infty}^2\sigma_3^2 \, , \label{asybe}
\ee
where $\sigma_3 = d\psi +\cos\theta\, d\phi$, and $L_{\infty} = \sqrt{\rho_0(\rho_0 +\rho_1)}$ is
the radius of the compact extra dimension. \textcolor{black}{The above asymptotic spacetime is locally
asymptotically flat and has a boundary $R \times S^1 \hookrightarrow S^2$, which is the same one as
that of the KK magnetic monopole. Concretely speaking, it is a metric of a `twisted' $S^1$ fiber bundle
over a four-dimensional Minkowski spacetime, from which it is seen that the $S^1$ circle parameterized
by a coordinate $\psi$ has finite size even at the spatial infinity. The non-trivial twisting of the
$S^1$ circle fibred over the $S^2$ base space leads a four-dimensional U(1) gauge field by KK reduction
under which the length $L_{\infty}$ is usually related to the NUT charge.}

The static squashed three-charge black hole solution has already been given in Ref. \cite{PLB726-404},
exactly with the above asymptotic behavior (\ref{asybe}). Its metric and the three Abelian gauge
potentials are expressed as follows:
\bea\label{3csquashed}
ds^2 &=& (H_1H_2H_3)^{1/3}\bigg[-\frac{1 -\rho_1/\rho}{H_1H_2H_3}d\tau^2
 +\frac{1 +\rho_0/\rho}{1 -\rho_1/\rho}d\rho^2 \nn \\
&& +\rho(\rho +\rho_0)(d\theta^2 +\sin^2\theta d\psi^2) \nn \\
&&\quad +\frac{\rho_0(\rho_0 +\rho_1)}{1 +\rho_0/\rho}(d\phi +\cos\theta d\psi)^2\bigg] \, , \\
&&\hspace*{-25pt}\textcolor{black}{\phi_1 = \frac{1}{\sqrt{6}}\ln\Big(\frac{H_3^2}{H_1H_2}\Big) \, ,
\quad \phi_2 = \frac{1}{\sqrt{2}}\ln\Big(\frac{H_2}{H_1}\Big) \, , } \\
A_I &=& \frac{c_Is_I\rho_1}{H_I\rho}d\tau \, ,
\eea
where $c_I = \cosh\delta_I$, and $s_I = \sinh\delta_I$, and three scalar functions are
\bea\label{3css}
X_I = \frac{(H_1H_2H_3)^{1/3}}{H_I}\, ,  \qquad H_I = 1 +s_I^2\frac{\rho_1}{\rho} \, .
\eea
Note that the above solution can be obtained via a solution-generating technique from its uncharged
version where all three charge parameters are set to zero. \textcolor{black}{Intriguingly, such static
three-charge squashed KK spacetime looks fully five-dimensional in the vicinity (namely, near-horizon
region) of the black hole, while (asymptotically) it resembles a four-dimensional object with a compact
fifth dimension at infinity.}

Incidently, we just delivery the supersymmetric BPS solutions for the three-charge static squashed
black hole within the $STU$ model:
\bea
ds^2 &=& \big(H_1H_2H_3\big)^{1/3}\bigg[-\frac{d{\tau}^2}{H_1H_2H_3}
 +\frac{\rho+\rho_0}{\rho}d{\rho}^2 \nn \\
&& +\rho(\rho +\rho_0)(d{\theta}^2 +\sin^2{\theta}d{\phi}^2) \nn \\
&&\quad +\frac{\rho_0^2\rho}{\rho +\rho_0}(d{\psi} +\cos{\theta}d{\phi})^2 \bigg] \, , \\
A_I &=& \pm\frac{q_I}{H_I\rho}d{\tau}\, ,  \quad
\eea
in which $H_I = 1 +q_I/\rho$, for $I = 1,2,3$.

\section{Four-charge static squashed black hole solution}\label{III}

In this section, we introduce the $STU-W^2U$ model of five-dimensional $\cN = 2$ supergravity
theory underpinning our squashed black hole solution. This model is specified by the pre-potential
\be
\cV = \frac{1}{6}C_{IJK}X^IX^JX^K = STU -W^2U = 1 \, ,
\ee
where the non-vanishing components of the symmetric tensor $C_{IJK}$ are: $C_{123}=1$ and
$C_{344} = C_{434} = C_{443} = -2$. By setting $X^4 = W = 0$, it reduces to the standard
$STU$ model.

The corresponding bosonic Lagrangian for the five-dimensional $\mathcal{N}=2$, $STU-W^2U$
ungauged supergravity coupled to three vector multiplets is given by\footnote{\textcolor{black}{
For details please refer to subsections (2.1), (3.1) and (3.2) in Ref. \cite{JHEP0226252} for
the present model, and also especially Ref. \cite{NPB553-317} for the $STU$ model.}}:
\bea\label{Lnew}
S_5 &=& \frac{1}{16\pi G}\int d^{5}x\bigg\{
 \sqrt{-g}\Big[R -3(\partial\varphi_1)^2 -\alpha(\partial\varphi_2)^2 \nn \\
&& -\frac{(\partial\alpha)^2}{4\alpha(\alpha-1)}
 -\frac{\alpha}{4}\big(e^{-2\varphi_1-2\varphi_2}F_1^2 +e^{-2\varphi_1+2\varphi_2}F_2^2\big) \nn \\
&& -\frac{1}{4}e^{4\varphi_1}F_3^2 -\frac{2\alpha-1}{2}e^{-2\varphi_1}F_4^2
 -\frac{\alpha-1}{2}e^{-2\varphi_1}F_1F_2 \nn \\
&& +\sqrt{\alpha(\alpha-1)}\big(e^{-2\varphi_1-\varphi_2}F_1F_4
 +e^{-2\varphi_1 +\varphi_2}F_2F_4\big)\Big] \nn \\
&&\quad -\frac{1}{4}\varepsilon^{\mu\nu\alpha\beta\lambda}\big(F_{1\mu\nu}F_{2\alpha\beta}
 -F_{4\mu\nu}F_{4\alpha\beta}\big)A_{3\lambda} \bigg\} \, ,
\eea
where $F^I = dA^I$ are the field strength two-forms for the four $U(1)$ gauge potential one-forms $A^I$,
and the scalar fields $(\varphi_1, \varphi_2, \alpha)$ are related to the $X^I$'s by:
\be\begin{aligned}
X^1&= \sqrt{\alpha}\, e^{\varphi_1+\varphi_2} \equiv S \, ,
& X^2&= \sqrt{\alpha}\, e^{\varphi_1-\varphi_2} \equiv T \, , \\
X^3&= e^{-2\varphi_1} \equiv U \, ,
& X^4&= \sqrt{\alpha-1}\, e^{\varphi_1} \equiv W \, .
\end{aligned}\ee

With the above asymptotic behavior (\ref{asybe}) in hand, we have adopted a brute-force method
to seek a simplified expression for the static four-charge squashed black hole solution, whose
explicit expressions of the metric and the four Abelian gauge potentials are given below
\bea
ds^2 &=& \big(Z_1Z_2-Z_4^2\big)^{1/3}Z_3^{1/3}\bigg[-\frac{f(\rho)}{
 \big(Z_1Z_2-Z_4^2\big)Z_3}d{\tau}^2 \nn \\
&& +\frac{(\rho+\rho_0)}{\rho f(\rho)}d{\rho}^2
 +\rho(\rho +\rho_0)(d{\theta}^2 +\sin^2{\theta}d{\phi}^2) \nn \\
&&\quad +\frac{L_{\infty}^2\rho}{\rho +\rho_0}(d{\psi}
 +\cos{\theta}d{\phi})^2 \bigg] \, , \label{4csquashed} \\
&&\hspace*{-25pt}\textcolor{black}{\varphi_1 = \frac{1}{6}
 \ln\Big(\frac{Z_3^2}{Z_1Z_2 -Z_4^2}\Big) \, , \quad
\varphi_2 = \frac{1}{2}\ln\Big(\frac{Z_2}{Z_1}\Big) \, ,} \nn \\
&&\hspace*{-25pt} \textcolor{black}{\alpha = \frac{Z_1Z_2}{Z_1Z_2 -Z_4^2} \, , \quad {\textrm or}\quad
\varphi_3 = \ln\bigg(\frac{\sqrt{Z_1Z_2}+Z_4}{\sqrt{Z_1Z_2-Z_4^2}}\bigg)} \\
A_1 &=& \frac{p_1Z_2 -p_4Z_4}{\rho\big(Z_1Z_2-Z_4^2\big)}d{\tau}\, ,  \quad
A_2 = \frac{p_2Z_1 -p_4Z_4}{\rho\big(Z_1Z_2-Z_4^2\big)}d{\tau}\, ,    \nn \\
A_3 &=& \frac{p_3}{\rho Z_3}d{\tau}\, ,  \qquad \label{4csqp}
A_4 = \frac{q_4(p_2Z_1 -p_1Z_2)}{(q_1 -q_2)\rho(Z_1Z_2 -Z_4^2)}d{\tau}\, , \quad
\eea
in which
\bea
f(\rho) &=& 1 -\frac{2m}{\rho} +\frac{p_4^2 +(w-1)q_4^2}{\rho^2} \, , \nn
\eea
\bea
Z_I &=& 1 +\frac{q_I}{\rho}  \quad (I = 1,2,3) \, , \qquad Z_4 = \frac{q_4}{\rho} \, , \nn \\
p_1 &=& \sqrt{q_1^2 +2mq_1 +wq_4^2}\, , \qquad p_2 = \sqrt{q_2^2 +2mq_2 +wq_4^2} \, , \nn \\
p_3 &=& \sqrt{q_3^2 +2mq_3 +p_4^2 +(w-1)q_4^2}\, , \quad p_4 = q_4 \frac{p_1 -p_2}{q_1 -q_2} \, . \nn
\eea
When $q_4 = p_4 = 0$ with $q_I = 2ms_I^2$, $p_I = 2ms_I c_I$, and $Z_I = H_I = 1 +2ms_I^2/\rho$ for
$I = 1, 2, 3$, the above solution (\ref{4csquashed}) reduces directly to the three-charge squashed
black hole solution \cite{PLB726-404} as is given in the above Eqs. (\ref{3csquashed}-\ref{3css}).

The above solution is obviously much more simple than that is obtained via directly applying the
squashing transformation \textcolor{black}{(see footnote \ref{fn1})} since one needs to perform further
coordinate transformations and very complicated identification of the solution parameters. The simplicity
of the above solution greatly facilitates the computation of conserved charges via the counterterm
method, whose expressions are also very concise as will be obviously shown below.

Now we briefly describe the solving steps that lead to the above solution, which is essentially
the same manipulation as that was performed in its corresponding un-squashed case \cite{JHEP0226252}
\footnote{\textcolor{black}{Both two five-dimensional metrics are written in the ansatz of timelike
dimensional reduction: $ds_5^2 = \big(Z_1Z_2-Z_4^2\big)^{1/3}Z_3^{1/3}\big[-f\, dx_5^2/\big(Z_1Z_2 -Z_4^2
\big)/Z_3 +ds_4^2\big]$, the difference lies in the four-dimensional spatial parts.}}, with the merely
difference between them being their different asymptotic structures of the metrics and function
dependence on the radical coordinates: $\rho \leftrightarrow r^2$. First, considering the asymptotic
structure (\ref{asybe}) that the metric must posses and requiring that the solution must reduce
to the three-charge case, and also that the moduli of the three scalar fields vanish at infinity,
we can write down the metric ansatz given in Eq. (\ref{4csquashed}) with the general form for the
structure function: $f(\rho) = 1 -\rho_1/\rho +\rho_2/\rho^2$, and the related four scalar functions
$Z_I$'s. Next, we solve the field equations for four Abelian potential one-forms and demand that
their temporal components vanish at infinity (or via making proper gauge transformations to remove
their constant values at infinity). By adjusting the integral constants to abandon the possible
$arctanh$-dependent \textcolor{black}{or logarithmic} terms in the gauge one-forms, we finally arrive
at the elegant expressions for the four Abelian gauge potentials (\ref{4csqp}). In doing so, we see
that the fourth constant is not dependent, but can be rewritten as: $p_4 = q_4(p_1-p_2)/(q_1-q_2)$.
In the third step, we solve three scalar field equations and find that the necessary conditions for
their fulfilments are: $p_3^2 = q_3^2f(-q_3) = q_3^2 +\rho_1q_3 +\rho_2$, and
\bea
\rho_1 &=& \frac{p_1^2 -q_1^2 -p_2^2 +q_2^2}{q_1-q_2} \, , \nn \\
\rho_2 &=& p_4^2 -q_4^2 +\frac{q_1(p_2^2 -q_2^2) +q_2(q_1^2 -p_1^2)}{q_1-q_2} \, . \nn
\eea
Then, we verify the Einstein equation and find that all their components vanish after we let:
$L_{\infty}^2 = \rho_0^2f(-\rho_0) = \rho_0^2 +\rho_0\rho_1 +\rho_2$. At this step, all
equations of motion are completely satisfied.

As our final step, we shall re-parameterize the integral constants to relate them to the solution
parameters ($\rho_1=2m$ and $q_I$). By setting $\rho_1 = 2m$, we can obtain: $p_1^2 = q_1^2 +2mq_1
+wq_4^2$ and $p_2^2 = q_2^2 +2mq_2 +wq_4^2$, where $w$ is an arbitrary constant which can be set
to zero in order to recover their usual expressions in the $STU$ model. With this parametrization,
we find that $\rho_2 = p_4^2 +(w-1)q_4^2$, which can not be put to zero, in general, unless in the
$STU$ case or in the supersymmetric BPS case, \textcolor{black}{and or in an interesting case that will
be discussed in details below in the subsection \ref{BHsp}}. In the second case, we can simply let
the mass parameter vanish ($m = 0$), see the Appendix for the explicit solutions. The non-vanishing of
$\rho_2$ characterizes that no solution-generating technique can be available in the $STU-W^2U$ model,
unlike that in the $STU$ model. Note also that we just take the plus sign for $p_I$ for $I=1, 2, 3$
in the final expressions.

\textcolor{black}{
\subsection{Main steps to solve the field equations}
All field equations and related definitions associated to the Lagrangian (\ref{Lnew}) have already
been displayed in the subsection (3.2) of Ref. \cite{JHEP0226252}. For the four harmonic functions:
$Z_I(\rho) = h_I +q_I/\rho$, we will take the convenient choice: $h_I = (1, 1, 1, 0)$ for $I=1,2,3,4$
so that the modules of all three scalar fields ($\varphi_1, \varphi_2, \varphi_3$) vanish
asymptotically at infinity, where we have redefined $\alpha
= \cosh^2\varphi_3$.}

\textcolor{black}{We begin with by solving the modified Maxwell field equations for the four Abelian
gauge one-forms: $A^I = K_I(\rho)d\tau$. First, the equation of the third Abelian vector field can
be expressed as:
\be
\p_{\rho} \big(\rho^2Z_3^2\p_{\rho}\, K_3\big) = 0 \, .
\ee
Integrating out it once, we can obtain: $\rho^2Z_3^2\p_{\rho}\, K_3 = -p_3$, in which the integration
constant is proportional to the electric charge of this gauge field; integrating once more and
abandoning a redundant constant which can be gauged away by using the gauge transformation, we
therefore arrive at: $K_3 = p_3/(\rho +q_3)$.}

\textcolor{black}{The remaining three Abelian field equations can be collectively written as:
\bea
&& \p_{\rho}\big[\rho^2\big(Z_1^2\p_{\rho}\, K_1 +Z_4^2\p_{\rho}\, K_2
 -2Z_1Z_4\p_{\rho}\, K_4\big)\big] = 0 \, , \nn \\
&& \p_{\rho}\big[\rho^2\big(Z_2^2\p_{\rho}\, K_2 +Z_4^2\p_{\rho}\, K_1
 -2Z_2Z_4\p_{\rho}\, K_4\big)\big] = 0 \, , \\
&& \p_{\rho}\big[\rho^2\big(Z_1Z_4\p_{\rho}\, K_1 +Z_2Z_4\p_{\rho}\, K_2
 -(Z_1Z_2 +Z_4^2)\p_{\rho}\, K_4\big)\big] = 0 \, . \nn
\eea
One can similarly arrange their three integration constants as $(-p_1, -p_2, -p_4)$, and rewrite
them as:
\bea
\p_{\rho}\, K_1 &=& -\frac{p_1Z_2^2 +p_2Z_4^2 -2p_4Z_2Z_4}{\rho^2
 \big(Z_1Z_2-Z_4^2\big)^2} \, , \nn \\
\p_{\rho}\, K_2 &=& -\frac{p_2Z_1^2 +p_1Z_4^2 -2p_4Z_1Z_4)}{\rho^2
 \big(Z_1Z_2-Z_4^2\big)^2} \, , \\
\p_{\rho}\, K_4 &=& -\frac{p_1Z_2Z_4 +p_2Z_1Z_4
 -p_4\big(Z_1Z_2 +Z_4^2\big)}{\rho^2\big(Z_1Z_2-Z_4^2\big)^2} \, , \nn
\eea
which can be apparently integrated out as
\bea
K_1 &=& \frac{p_1(\rho+q_2) -p_4q_4}{(\rho +q_1)(\rho+q_2) -q_4^2} \nn \\
 &&\qquad -q_4\int\frac{p_4(q_1-q_2) -q_4(p_1-p_2)}{\big[(\rho +q_1)(\rho+q_2)
  -q_4^2\big]^2}d\rho \, , \nn \\
K_2 &=& \frac{p_2(\rho+q_1) -p_4q_4}{(\rho +q_1)(\rho+q_2) -q_4^2} \nn \\
 &&\qquad +q_4\int\frac{p_4(q_1-q_2) -q_4(p_1-p_2)}{\big[(\rho +q_1)(\rho+q_2)
  -q_4^2\big]^2}d\rho \, , \\
K_4 &=& \frac{q_4}{q_1-q_2}\times \frac{p_2(\rho+q_1)
 -p_1(\rho+q_2)}{(\rho +q_1)(\rho+q_2) -q_4^2} \nn \\
&& +\Big(p_4 -q_4\frac{p_1-p_2}{q_1-q_2}\Big)
\int\frac{(\rho +q_1)(\rho+q_2) +q_4^2}{\big[(\rho +q_1)(\rho+q_2)
 -q_4^2\big]^2}d\rho \, , \nn
\eea
in each of these three expressions, the last term that has to be integrated out involves the
\textcolor{black}{$arctanh$ or logarithmic} functions plus another rational functions of $\rho$
that can be disregarded after imposing the constraint condition: $p_4 = q_4(p_1 -p_2)/(q_1 -q_2)$.
In the meanwhile, the three integration constants have been also set to zero for the reason
explained in the above.
}

\textcolor{black}{Next, we turn to solve two core ingredients for the resulting Einstein field
equations written in the mixed form: ($E^{\mu}_{~\nu} = 0$). In particular, two compositions
of Einstein's equations: $E^{\rho}_{~\rho} +E^{\theta}_{~\theta} = 0$ and $E^{\psi}_{~\psi}
 -E^{\phi}_{~\phi} = 0$ become proportional to
\bea
&& \rho^2\p^2_{\rho}\, f +4\rho\p_{\rho}\, f +2f -2 = 0 \, , \\
&& \rho^2(\rho +\rho_0)\p_{\rho}\, f +\rho(\rho +2\rho_0)f \nn \\
&&\qquad\qquad -(\rho +\rho_0)^2 +L_{\infty}^2 = 0\, , \qquad
\eea
The integration of them by using the `\emph{dsolve}' tool within the Maple package, give rise
to two different expressions for $f(\rho)$, respectively,
\bea
f &=& \frac{\rho^2 -\rho_1\rho +\rho_2}{\rho^2} \, , \\
f &=& \frac{(\rho -C)(\rho +\rho_0) +L_{\infty}^2}{\rho^2} \, ,
\eea
in which $\rho_1, \rho_2$ and $C$ are three integration constants. The consistency between them
boils down to the following two conditions:
\be
C = \rho_1 +\rho_0\, , \quad L_{\infty}^2 = \rho_0^2 +\rho_1\rho_0 +\rho_2 \, .
\ee}

\textcolor{black}{Subsequently, both equations for the scalar field $\varphi_2$ and $\alpha =
\cosh^2\varphi_3$ reduce to
\bea
&& \rho^2\big(Z_1Z_2) -Z_4^2\big)\p_{\rho}f +\big(q_1Z_2 +q_2Z_1 -2q_4Z_4\big)f \nn \\
&&\qquad -\frac{p_1Z_2 -p_2Z_1}{q_1-q_2} \big(p_1Z_2 +p_2Z_1 -2p_4Z_4\big) = 0 \, ,
\eea
into which with the substitution $f = 1 -\rho_1/\rho +\rho_2/\rho^2$, we will have
}

\textcolor{black}{
\bea
\rho_1 &=& \frac{p_1^2 -p_2^2}{q_1-q_2} -q_1 -q_2 \, , \nn \\
\rho_2 &=& p_4^2 -q_4^2 +q_1q_2 +\frac{q_1p_2^2-q_2p_1^2}{q_1-q_2} \, . \nn
\eea
}

\textcolor{black}{Finally, with all the expressions in hand, both the equation for the scalar
field $\varphi_1$ and the remaining five diagonal parts of the Einstein equations ($E^{\mu}_{~\nu}
= 0$) can be satisfied by setting $p_3^2 = q_3(q_3 +\rho_1) +\rho_2$. This completes our solving
procedure.}

\textcolor{black}{The above brute-force solving strategy completely applies to the case with the general
harmonic ansatz for the scalar fields: $Z_I = h_I +q_I/\rho$. The asymptotic expansions of three scalar
fields near the infinity read
\be
\varphi_i = \varphi_i^{\infty} +\frac{\Sigma_i}{\rho} +O(\rho^{-2}) \qquad (i=1,2,3) \, ,
\ee
where their moduli and scalar charges are, respectively,
\bea
\varphi_1^{\infty} &=& \frac{1}{6}\ln\Big(\frac{h_3^2}{h_1h_2 -h_4^2}\Big) \, , \qquad
\varphi_2^{\infty} = \frac{1}{2}\ln\Big(\frac{h_2}{h_1}\Big) \, , \nn \\
\varphi_3^{\infty} &=& \ln\big(\sqrt{h_1h_2} +h_4\big) -\frac{1}{2}\ln\big(h_1h_2 -h_4^2\big) \, , \nn \\
\Sigma_1 &=& \frac{q_3}{3h_3} -\frac{q_1h_2 +q_2h_1 -2q_4h_4}{6\big(h_1h_2 -h_4^2\big)} \, , \quad
\Sigma_2 = \frac{q_2h_1 -q_1h_2}{2h_1h_2} \, , \nn \\
\Sigma_3 &=& \frac{2q_4h_1h_2 -(q_1h_2 +q_2h_1)h_4}{2\sqrt{h_1h_2}\big(h_1h_2 -h_4^2\big)} \, , \nn
\eea
which will directly enter into the first law and integral mass formula \cite{PRL77-4992}. Besides these,
as the reason will be immediately explained below, the arbitrary values of $h_I$'s will also indirectly
attribute to the mass formula via the expressions of other thermodynamical quantities. Clearly three moduli
vanish in the convenient choice: $h_I = (1,1,1,0)$ and make no contribution into the first law and the
Bekenstein-Smarr formula, namely, when the limit of the moduli to our preferred choice: $h_I = (1, 1, 1, 0)$
is being taken, all the final results recover our present simplest ones.
}

\textcolor{black}{On the other hand, the structure function takes the general quadratic form: $f =
f_0 -f_1/\rho +f_2/\rho^2$, in which their three coefficients are very involved, especially, the
expression $f_0$ is composed of $h_I$ and $q_I$, which does not approach to unity at infinity. In
the same time, the extra length $L_{\infty}$ acquires a very complicated expression also. Since $f_0
\not= 1$ and $h_I$'s are arbitrary constant, in order to acquire the asymptotic metric (\ref{asybe}),
one has to rescale the radial and temporal coordinates, so that not only the mass and stress
tension, but also the entropy and the temperature own very cumbersome expressions, although they can
recover their concise forms in the simplest gauge: $h_I = (1,1,1,0)$, which is the most convenient
choice for us to find a much simpler solution at the very beginning. What is more, the four Maxwell
fields will not be the simplest ones as those given in the above, with the resulting expressions for
the four electric charges and electrostatic potentials being very tedious also. Both these considerations
constitute our motivation to seek a simpler metric for the squashed black hole in the ungauged supergravity
theory.
}

\textcolor{black}{
\subsection{Black hole splitting effect induced by the fourth vector field}\label{BHsp}
Note that if the parameter $w$ introduced in the above is arbitrarily variable, which means that
$\rho_2 = p_4^2 +(w-1)q_4^2$ does not vanish in general, then this would imply that there may be
no solution generating technique to derive this solution from the vacuum seed squashed Schwarzschild
metric. In the meanwhile, the electric charge of the third Abelian vector field does not vanish
even when the third charge parameter reduces to zero ($q_3 = 0$). On the other hand, there exists
another one promising possibility for which one can let $\rho_2 = 0$ by choosing $w$ to satisfy a
certain identity. In the general case where $q_2\not = q_1$ and $q_4 \not= 0$, by inserting $p_4
= q_4(p_1 -p_2)/(q_1 -q_2)$ into the expressions for $\rho_2 = p_4^2 +(w-1) q_4^2$, this condition
becomes a quadratic equation for $w$:
\bea
&& \Big[q_4^2 +\frac{1}{4}(q_1-q_2)^2\Big]w^2 +\big[(q_1+q_2)m \nn \\
&&\qquad\qquad +q_1q_2-q_4^2\big]w +m^2 = 0 \, , \label{ww}
\eea
which generally has two distinct roots expressed in term of four solution parameters ($m, q_1, q_2,
q_4$):
\bea
w &=& -\,\frac{m(q_1+q_2) +q_1q_2-q_4^2}{2\big[q_4^2 +(q_1-q_2)^2/4\big]} \nn \\
&& \pm\frac{\sqrt{q_1q_2-q_4^2}\sqrt{(q_1+2m)(q_2+2m)-q_4^2}}{2\big[q_4^2 +(q_1-q_2)^2/4\big]} \, .
\eea
Since the above solution degenerates to that given in Ref. \cite{PLB726-404} of the usual $STU$ model
when $q_4 = 0$. The fact that $w$ has two distinct roots demonstrates that with the addition of the
nonzero fourth charge of the extra Abelian gauge field, the solution of the STU model is split into
two different branches. One can name this novel phenomenon as the black hole splitting effect induced
by the Maxwell field, somewhat like the Stark effect -- atomic spectrum lines splitting in the
external electric field in the atom physics. To the best of the authors' knowledge, this feature has
never been reported by any one  before in the previous researches of black hole physics. The above
conclusion is also in accordance with the BPS case which is present in the Appendix, that corresponds
to the limiting case $\rho_1 = 2m \to 0$ and $\rho_2 \to 0$.}

\textcolor{black}{Note also that when $\rho_2 = 0$, then $p_3 = \sqrt{q_3(q_3+2m)}$ and when $q_3 = 0$,
and then the electric charge of the third gauge field vanishes exactly as in the $STU$ model. The
special case where $q_2 = q_1$ will be separately treated in details in the next section.}

\textcolor{black}{
\section{Special case: $S^2U-W^2U$model}\label{IV}
%
In the section \ref{III}, it is pointed out that in the case when $q_4 = 0$ and $Z_4 = 0$, the
$STU-W^2U$ model trivially reduces to the familiar $STU$ model. Here, we shall show that an
interesting special case when $q_2 = q_1$ and $Z_2 = Z_1$, the model degenerates to the $S^2U-W^2U$
model that can be associated with the $\tilde{S}\tilde{T}U$ model also. In this special case, the
pre-potential becomes $\cV = S^2U -W^2U = 1$, because $T=S$. One can set: $S+W=\tilde{S}$ and $S-W
=\tilde{T}$, so that it reduces to that of the standard $\tilde{S}\tilde{T}U$ model $\cV = \tilde{S}
\tilde{T}U = 1$.
}

\textcolor{black}{Next, because $\varphi_2 = 0$ and $F_2 = F_1$ (and $A_2 = A_1$), then after setting
$\alpha = \cosh^2\varphi_3$, the Lagrangian (\ref{Lnew}) degenerates to
\bea
S_5 &=& \frac{1}{16\pi G}\int d^{5}x\bigg\{
 \sqrt{-g}\Big[R -3(\partial\varphi_1)^2 -(\partial\varphi_3)^2
  -\frac{1}{4}e^{4\varphi_1}F_3^2 \nn \\
&& -\frac{1}{2}e^{-2\varphi_1}\cosh(2\varphi_3) \big(F_1^2 +F_4^2\big) 
 +\sinh(2\varphi_3) e^{-2\varphi_1}F_1F_4\Big] \nn \\
&&\quad -\frac{1}{4}\varepsilon^{\mu\nu\alpha\beta\lambda}\big(F_{1\mu\nu}F_{1\alpha\beta}
 -F_{4\mu\nu}F_{4\alpha\beta}\big)A_{3\lambda} \bigg\} \, ,
\eea
which is equivalent to that of the $\tilde{S}\tilde{T}U$ model given by Eq. (\ref{L-STU}) with the
identifications: $F_1 +F_4 = \tilde{F_1}$ and $F_1 -F_4 = \tilde{F_2}$ for the Maxwell fields (and
then omitting the tildes), $\varphi_1 = \phi_1/\sqrt{6}$ and $\varphi_3 = \phi_2/\sqrt{2}$ for the
scalar fields. This replacements can also be confirmed at the level of the equations of motion.
}

\textcolor{black}{On the other hand, at the level of the solution corresponding to the above Lagrangian
action, we can solve to obtain
\bea
ds^2 &=& \big(Z_1^2-Z_4^2\big)^{1/3}Z_3^{1/3}\bigg[-\frac{f(\rho)}{
 \big(Z_1^2-Z_4^2\big)Z_3}d{\tau}^2 \nn \\
&& +\frac{(\rho+\rho_0)}{\rho f(\rho)}d{\rho}^2
 +\rho(\rho +\rho_0)(d{\theta}^2 +\sin^2{\theta}d{\phi}^2) \nn \\
&&\quad +\frac{L_{\infty}^2\rho}{\rho +\rho_0}(d{\psi}
 +\cos{\theta}d{\phi})^2 \bigg] \, , \\
&&\hspace*{-25pt} \textcolor{black}{\varphi_1 = \frac{1}{6}
 \ln\Big(\frac{Z_3^2}{Z_1^2 -Z_4^2}\Big) \, ,
\quad \varphi_3 = \frac{1}{2}\ln\Big(\frac{Z_1 +Z_4}{Z_1 -Z_4}\Big) \, ,} \\
A_1 &=& \frac{p_1Z_1 -p_4Z_4}{\rho\big(Z_1^2-Z_4^2\big)}d{\tau}\, , \quad
A_3 = \frac{p_3}{\rho Z_3}d{\tau}\, , \nn \\
A_4 &=& \frac{p_4Z_1 -p_1Z_4}{\rho(Z_1^2 -Z_4^2)}d{\tau}\, ,
\eea
in which $Z_1 = 1 +q_1/\rho$, $Z_3 = 1 +q_3/\rho$, $Z_4 = q_4/\rho$, and
\bea
f(\rho) = 1 -\frac{f_1}{\rho} +\frac{f_2}{\rho^2} \, ,
\eea
where
\bea
f_1 = \frac{2p_1p_4}{q_4} -2q_1 \, , \quad
f_2 = p_4^2 -q_4^2 +p_1^2 -q_1^2 -2q_1f_1 \, . \nn
\eea
This solution can also be obtained from the general solution (before parameterizations) given
in the last section by letting:
\bea
q_2 = q_1 +\epsilon \, \, \qquad p_2 = p_1 +\frac{p_4}{q_4}\epsilon \, , \nn
\eea
and taking the $\epsilon\to 0$ limit, thus one has $Z_2\to Z_1$ and $A_2\to A_1$.
}
\vskip -6pt
\textcolor{black}{In order to identify the above solution with that is given in Eqs. (\ref{3csquashed}-\ref{3css})
for the $STU$ model, we first note that we must let $Z_1 +Z_4 = H_1$, $Z_1 -Z_4 = H_2$ and $H_3 = Z_3$, so
that $A_1 +A_4 = \tilde{A_1}$ and $A_1 -A_4 = \tilde{A_2}$. In turn, these give rise to
\bea
&& q_1 +q_4 = \rho_1s_1^2 \, , \qquad q_1 -q_4 = \rho_1s_2^2 \, , \qquad q_3 = \rho_1s_3^2 \, , \nn \\
&& p_1 +p_4 = \rho_1c_1s_1 \, , \quad p_1 -p_4 = \rho_1c_2s_2 \, , \quad p_3 = \rho_1c_3s_3 \, . \nn
\eea
Substituting all these relations into the expressions of $f_1$ and $f_2$ given in the above, we
then have $f_1 = \rho_1$ and $\rho_2 = 0$.
}

\textcolor{black}{We close up this section by pointing out that the important significance of the
special case $q_2 = q_1$ studied in the above is to hint that perhaps one should let $f_2 = 0$
in the most general $STU-W^2U$ model also.
}

\section{Thermodynamics}\label{V}

\textcolor{black}{The thermodynamic properties of various squashed KK black holes had been widely
investigated in a lot of literatures, for instance, \cite{PRD85-064021,EPJC73-2377,PRD78-124006,
CQG24-4525,CQG25-085006,IJMPA24-2357,PRD84-124040}. In particular, the Euclidean action and free
energy was calculated in Refs. \cite{CQG24-4525,CQG25-085006} by using the standard thermodynamic
method.}

\textcolor{black}{Now let's turn to investigate its thermodynamic properties of our static squashed
solution.} The counterterm method \cite{PLB634-531,PRD73-044014} is well-suited for this task, as
it allows for the computation of conserved charges in our asymptotically locally flat spacetime with
boundary topology $R \times S^1 \hookrightarrow S^2$, \textcolor{black}{the same one as the Kaluza-Klein
magnetic monopole}. This method regularizes the gravitational action by adding to the boundary term
at infinity a counterterm that depends only on curvature invariants of the induced metric, leaving
the equations of motion unchanged.

To our end, we adopt the Gibbons-Hawking boundary term and the counterterm suggested by Mann and
Stelea \cite{PLB634-531} to regulate the potential divergence as follows:
\bea
I_b +I_{ct} &=& \frac{1}{8\pi G}\int\limits_{\p\cM} d^{4}x \sqrt{-h}K \nn \\
&& +\frac{1}{8\pi G}\int d^4x\sqrt{-h}\sqrt{2\cR}\, ,
\eea
In the boundary term, $K$ is the trace of extrinsic curvature $K_{ij}=(n_{i;j} +n_{j;i})/2$ for
the boundary $\p\cM$ with the induced metric $h_{ij}$, where $\mathcal{R}$ is the Ricci scalar
of the boundary metric $h_{ij}$. Variation of the above supplemented action with respect to
$h_{ij}$ yields the following boundary stress-energy tensor:
\be
8\pi T_{ij} = K_{ij}-h_{ij}K -\Psi(\cR_{ij}-h_{ij}\cR)
 -h_{ij}h^{kl}\Psi_{;kl}+\Psi_{;ij} \, ,
\ee
where $\Psi = \sqrt{2/\cR}$, and the covariant derivative in the above formulas is defined with
respect to the induced metric $h_{ij}$ on the boundary.

The counterterm mass and the nonvanishing gravitational tension can now be evaluated from the
following formulas:
\bea
M_{ct} &=& \frac{-1}{8\pi}\int_0^{2\pi}d\phi \int_0^{4\pi}d\psi \int_0^\pi d\theta\,
 \big(\!\sqrt{\Sigma}T^t_{~~t}\big)\big|_{\rho\to\infty} \, , \\
\cT &=& \frac{-1}{8\pi}\int_0^{2\pi}d\phi \int_0^\pi d\theta\,
 \big(\!\sqrt{\sigma}T^\psi_{~~\psi}\big)\big|_{\rho\to\infty} \, ,
\eea
where
\bea
&&\sqrt{\Sigma} = L_{\infty}\rho^{3/2}Z_3^{1/2}\sin\theta\sqrt{\rho +\rho_0}
 \sqrt{Z_1Z_2 -Z_4^2} \, , \nn \\
&&\sqrt{\sigma} = \big(Z_1Z_2 -Z_4^2\big)^{1/3}Z_3^{1/3}\rho(\rho +\rho_0)\sin\theta \, . \nn
\eea

Omitting the details of tedious computation, the counterterm mass and the non-vanishing
gravitational tension are rather concise and are given by
\bea
M_{ct} &=& \pi L_{\infty}(4m +q_1 +q_2 +q_3 +\rho_0)\, ,  \\
\cT &=& \frac{1}{2}(m +\rho_0) \, ,
\eea
where $L_{\infty} = \sqrt{\rho_0^2 +2m\rho_0 +\rho_2}$. \textcolor{black}{It had been shown in Refs.
\cite{PLB639-354,CQG25-085006} that the counterterm mass is identical to the Abbott-Deser mass of
the Ishihara-Matsuno squashed black hole evaluated on the flat reference background. In other words,
for any squashed KK black hole, the counterterm method is equivalent to the background subtraction
method using the flat background as a reference spacetime. This argument also applies to the present
solution.}

On the event horizon ($\rho_+ = m +\sqrt{m^2 -\rho_2}$), the Bekenstein-Hawking entropy $S = A/4$ and
the Hawking temperature $T = \kappa/(2\pi)$ are given by
\bea
S &=& 4\pi^2L_{\infty}\sqrt{Z_1Z_2 -Z_4^2}\sqrt{Z_3}\sqrt{\rho_+
 +\rho_0}\rho_+^{3/2}\Big|_{\rho = \rho_+} \, , \\
T &=& \frac{\p_{\rho} f(\rho)}{4\pi\sqrt{1 +\rho_0/\rho}
\sqrt{Z_1Z_2 -Z_4^2}\sqrt{Z_3}}\Big|_{\rho = \rho_+} \, ,
\eea
\textcolor{black}{where the entropy is consistent with Wald's formula. It would be anticipated to re-derive
this result by using the formal formulation advocated in Ref. \cite{JHEP0823039}.}

The electrostatic potentials $\Phi_I$ conjugate to the electric charges are defined at the
horizon by $\Phi_I = A_t^{I}(\rho_+)$ due to our convenient gauge choice and are given by
\be\begin{aligned}
&\Phi_1 = \frac{(p_1Z_2 -p_4Z_4)}{\rho(Z_1Z_2 -Z_4^2)}\Big|_{\rho=\rho_+} \, , ~~
\Phi_2 = \frac{(p_2Z_1 -p_4Z_4)}{\rho(Z_1Z_2 -Z_4^2)}\Big|_{\rho=\rho_+} \, ,  \\
&\Phi_3 = \frac{p_3}{\rho Z_3}\Big|_{\rho=\rho_+} \, , ~~
\Phi_4 = \frac{q_4(p_2Z_1 -p_1Z_2)}{(q_1-q_2)\rho(Z_1Z_2 -Z_4^2)}\Big|_{\rho=\rho_+} \, .
\end{aligned}\ee
The four electric charges associated with the gauge potentials are obtained by evaluating the
Gauss's integral
\be
Q_I = \pi\, L_{\infty} p_I \quad (I = 1,2,3,4) \, .
\ee

Finally, it can be shown that the above thermodynamic quantities simultaneously satisfy the
first law of black hole thermodynamics and the Bekenstein-Smarr mass formula:
\bea
dM_{ct} &=& TdS +\Phi_1dQ_1 + \Phi_2dQ_2 +\Phi_3dQ_3 -2\Phi_4dQ_4 \nn \\
&& +4\pi\mathcal{T}dL_{\infty} \, , \\
M_{ct} &=& \frac{3}{2}TS +\Phi_1Q_1 +\Phi_2Q_2 +\Phi_3Q_3 -2\Phi_4Q_4 \nn \\
&& +2\pi\mathcal{T}L_{\infty} \, ,
\eea
where $2\pi L_{\infty}$ is the length of the extra dimension, which is treated as a thermodynamic
variable to ensure the consistency of the Bekenstein-Smarr mass formula. \textcolor{black}{In the
viewpoint when the spacetime is reduced down to four dimensions, this length scale is usually
interpreted as the NUT charge after the spatial direction $\psi$ is Wick-rotated to a temporal
coordinate.}

\textcolor{black}{Note also that in the above we have treated $w$ as a genuine constant in order to
verify the first law and the integral mass formula. On the other hand, if one would like to adopt
the alternate perspective to parameterize the solution so that $\rho_2 = 0$. Then in this time,
one just has only one single horizon at $\rho = \rho_1 \equiv 2m$, and $w$ must be also viewed as
a true variable that is constrained by Eq. (\ref{ww}). Only after this constrain condition has been
taken into account, can the above mass formulae become consistent once again. Otherwise, they are
inconsistent any more.
}

\section{Conclusions}\label{VI}

In this work, we have obtained a novel form of the static four-charge squashed black hole solution
within the $STU-W^2U$ model of five-dimensional $\mathcal{N} = 2$ ungauged supergravity. We have sought
a fairly simpler expression for the solution so that the resulting thermodynamic quantities are also
very concise. Using the counterterm method, we then have computed the complete set of thermodynamic
quantities and demonstrated that they satisfy both the differential and integral mass formulae of
black hole thermodynamics, provided the extra-dimensional length is treated as a thermodynamic
variable.

\textcolor{black}{We have, for the first time, uncovered a novel black hole splitting phenomenon induced
by the extra charge -- namely, when the fourth Abelian vector field is added, the black hole solution
to the usual $STU$ model divides into two different branches in the new $STU-W^2U$ model. This new
observation indicates that the solution space of the $STU-W^2U$ model has a much richer structure than
that of the familiar $STU$ model. We also demonstrate that the special $S^2U-W^2U$ model can be attributed
to the usual $STU$ model which strongly suggests that one should set $\rho_2 = 0$ in the general
expression of the structure function: $f(\rho) = 1 -\rho_1/\rho +\rho_2/\rho^2$.}

A direct task in the future is to extend this work to consider the \textcolor{black}{rotating case
which corresponds to the cohomogeneity-one black hole}. Another interesting direction is to
investigate the supersymmetric BPS version as presented in the appendix. \textcolor{black}{Still there
is also a challenging possibility to embed the cohomogeneity-two Rasheed-Larsen black hole solution
\cite{NPB454-379,NPB575-211} into the present $STU-W^2U$ model like those that had already been
done in the $STU$ model \cite{PRD79-064020,CQG26-145006,PRD86-024022}.}

\medskip
\textbf{Acknowledgements}\quad
\textcolor{black}{We thank the anonymous referee for valuable comments and suggestions.} This work is
supported by the National Natural Science Foundation of China (NSFC) under Grants No. 12205243 and
No. 12375053, and by the Sichuan Science and Technology Program under Grant no. 2026NSFSC0021.

\medskip
\textbf{Data Availability Statement}\quad This manuscript has no associated data.
[Authors' comment: Data sharing not applicable to this article as no datasets were generated or
analysed during the current study.]

\medskip
\textbf{Code Availability Statement}\quad This manuscript has no associated code/software.
[Authors' comment: Code/Software sharing not applicable to this article as no code/software was
generated or analysed during the current study.]

\medskip
\textbf{Open Access}\quad
\medskip Funded by SCOAP$^3$.

\section*{Appendix: Supersymmetric four-charge static squashed black hole solutions}
\setcounter{equation}{0}
\renewcommand{\theequation}{A\arabic{equation}}

In this appendix, we present the supersymmetric BPS solutions for the four-charge static squashed
black hole within the $STU-W^2U$ model:
\bea
ds^2 &=& \big(Z_1Z_2-Z_4^2\big)^{1/3}Z_3^{1/3}\bigg[-\frac{d{\tau}^2}{
 \big(Z_1Z_2-Z_4^2\big)Z_3} \nn \\
&& +\frac{\rho+\rho_0}{\rho}d{\rho}^2
 +\rho(\rho +\rho_0)(d{\theta}^2 +\sin^2{\theta}d{\phi}^2) \nn \\
&&\quad +\frac{\rho_0^2\rho}{\rho +\rho_0}(d{\psi}
 +\cos{\theta}d{\phi})^2 \bigg] \, , \\
A_1 &=& \pm \frac{w_1Z_2 -w_4Z_4}{\rho\big(Z_1Z_2-Z_4^2\big)}d{\tau}\, , \quad
A_2 = \pm \frac{w_2Z_1 -w_4Z_4}{\rho\big(Z_1Z_2-Z_4^2\big)}d{\tau}\, , \nn \\
A_3 &=& \frac{w_3}{\rho Z_3}d{\tau}\, , \quad
A_4 = \pm \frac{q_4(w_2Z_1 -w_1Z_2)}{(q_1 -q_2)\rho(Z_1Z_2 -Z_4^2)}d{\tau}\, , \qquad
\eea
in which $w_3^2 = p_3^2$, $w_4 = q_4(w_1 -w_2)/(q_1 -q_2)$, and
\bea
Z_I = 1 +\frac{q_I}{\rho}  \quad (I = 1,2,3) \, , \quad Z_4 = \frac{q_4}{\rho} \, . \nn
\eea

For $q_4\not= 0$, there are two different branches:
\be
w_1 = q_1 \, , \quad  w_2 = q_2 \, , \quad w_4 = q_4 \, ,
\ee
or
\bea
&& w_1 = \frac{q_1^2-q_1q_2 +2q_4^2}{\sqrt{(q_1-q_2)^2 +4q_4^2}} \, , \quad
w_2 = \frac{q_2^2-q_1q_2 +2q_4^2}{\sqrt{(q_1-q_2)^2 +4q_4^2}} \, , \nn \\
&& w_4 = \frac{q_4(q_1 +q_2)}{\sqrt{(q_1-q_2)^2 +4q_4^2}} \, .
\eea
It is obvious that the supersymmetric BPS three-charge solution presented in the end of Sec. \ref{II}
is now split into two different branches with the addition of the fourth electric charge.

\end{document}